\documentclass[aps,prl,twocolumn,superscriptaddress,showpacs]{revtex4-1}
\usepackage{amsmath,amssymb}
\usepackage{graphicx}
\usepackage{natbib}
\usepackage{color}

\begin{document}

\title{Coherent Virtual Absorption Based on Complex Zero Excitation for Ideal Light Capturing}

\author{Denis~G.~Baranov}
\email[]{denisb@chalmers.se}
\affiliation{Department of Physics, Chalmers University of Technology, 412 96 Gothenburg, Sweden}
\affiliation{Moscow Institute of Physics and Technology, Dolgoprudny 141700, Russia}
\affiliation{ITMO University, St.~Petersburg 197101, Russia}

\author{Alex~Krasnok}
\affiliation{Department of Electrical and Computer Engineering, The University of Texas at Austin, Austin, Texas 78712, USA}

\author{Andrea~Al\`u}
\affiliation{Department of Electrical and Computer Engineering, The University of Texas at Austin, Austin, Texas 78712, USA}

\begin{abstract}
Absorption of light is directly associated with dissipative processes in a material. In suitably tailored resonators, a specific level of dissipation can support coherent perfect absorption, the time-reversed analogue of lasing, which enables total absorption and zero scattering in open cavities. On the contrary, the scattering zeros of lossless objects strictly occur at complex frequencies. While usually considered non-physical due to their divergent response in time, these zeros play a crucial role in the overall scattering dispersion. Here, we introduce the concept of coherent virtual absorption, accessing these modes by temporally shaping the incident waveform. We show that engaging these complex zeros enables storing and releasing the electromagnetic energy at will within a lossless structure for arbitrary amounts of time, under the control of the impinging field. The effect is robust with respect to inevitable material dissipation and can be realized in systems with any number of input ports. The observed effect may have important implications for flexible control of light propagation and storage, low-energy memory, and optical modulation.
\end{abstract}

\maketitle

\section{Introduction}
Linearity in optics ensures that the response to any time-varying input of an open system can be described through the frequency dependent scattering operator (or scattering matrix $S$), a complex-valued analytical function that relates the amplitudes of the outgoing and incoming radiation channels~\cite{Haus,Fan2003,Bonod13}. For the full description of the optical response of a linear scatterer, it is sufficient to know the scattering operator at real frequencies. However, the complex frequency response provides interesting insights into its dispersion. In the complex frequency plane, the eigenvalues of $S$ have two sets of singularities, zeros and poles: the first correspond to solutions of Maxwell's equations without outgoing fields, i.e., perfectly absorbing, while the latter correspond to lasing modes, with a divergent response for finite excitation~\cite{Bonod13,Bonod13b}. The position of these singularities in the complex plane determines the scattering matrix at any frequency, based on Weierstrass factorization theorem~\cite{Bonod13,Bonod13b}. In a reciprocal lossless system, without loss, the poles and zeros of S eigenvalues always exist in pairs and their positions are related via complex conjugation. Specifically, poles are always located in the lower half-plane, while zeros are located in the upper half-plane, which correspond to exponential attenuation and divergence in time, respectively. The presence of losses breaks the time-reversal symmetry, and a suitable combination of geometrical design and a specific amount of material loss can push one of these zeros to the real axis, enabling coherent perfect absorption, i.e., the possibility of totally absorbing the monochromatic wave without any scattering~\cite{Chong2010,Zhang2012,Yoon2012,Ignatov2016,CPA-arxiv}. This phenomenon is the time-reversed process of lasing, in which the right amount of gain can push one of the poles of S eigenvalues to the real frequency axis, enabling self-oscillations. Coherent perfect absorption allows manipulation of light with light via real absorption instead of nonlinearity. However, this approach has a disadvantage of real absorption and, therefore, high thermal heating.

In this paper, we generalize the concept of coherent perfect absorption to fully lossless and passive systems (with no Ohmic losses nor gain), by engaging a zero in the complex frequency plane to induce coherent virtual absorption of light, and its release from the system on demand. Instead of adding loss to the system to push the complex zero towards the real axis, we tailor the temporal dependence of the incident field, such that its temporal profile matches the exponentially diverging perfectly absorbing mode associated with the complex zero over a finite interval. Interestingly, we show that during this transient excitation the scattering form the structure totally vanishes, \emph{as if the structure were perfectly absorbing}, despite the absence of material loss. However, when the exponentially growing incident field cuts off, it gives rise to the release of the energy stored in the system. This counterintuitive result may pave the way to low-energy, ultrafast, light-controlled storing and release of optical signals on demand.

\begin{figure*}[t]
\includegraphics[width=1.5\columnwidth]{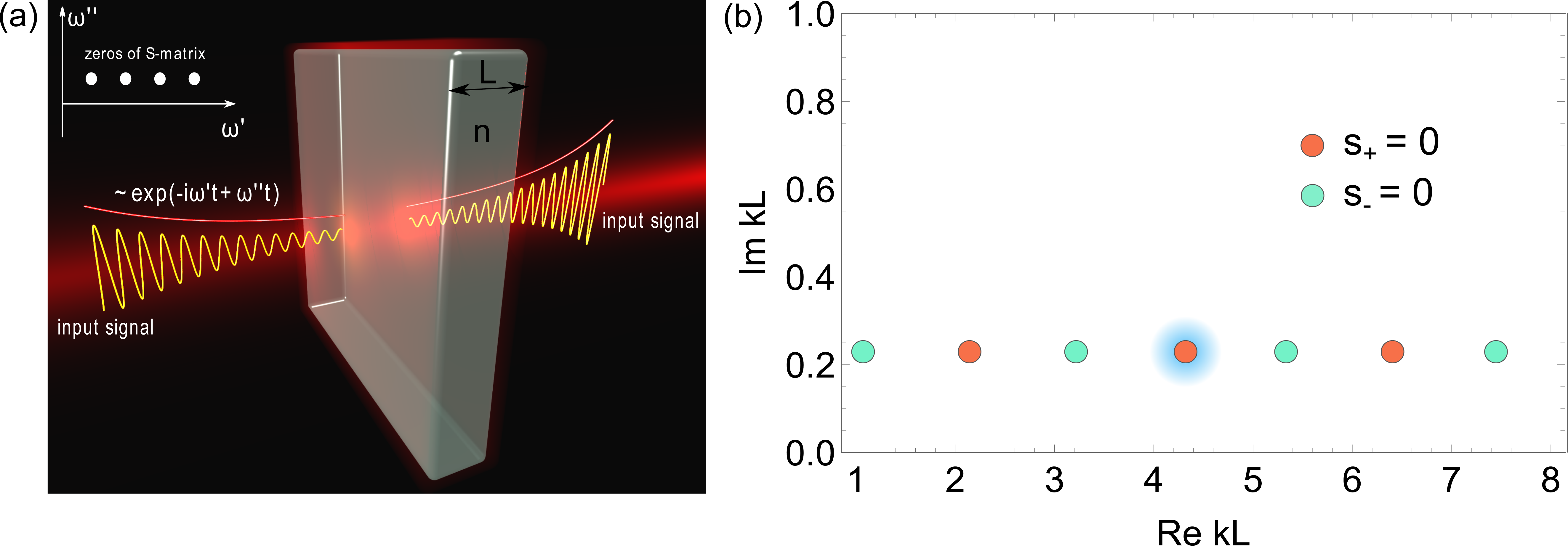}
\caption{(a) Schematic geometry of the problem. (b) Location of the $S$-matrix eigenvalues zeros of a non-magnetic planar slab with refractive index $n=3$ for normal incidence. Plus and minus signs refer to zeros of symmetric and anti-symmetric modes of the slab.}
\label{fig1}
\end{figure*}

\section{Results and Discussion}
Under an ${e^{ - i\omega t}}$ time convention, an arbitrary lossless object can have scattering zeros only at complex frequencies ${\omega _0} = \omega ' + i\omega ''$ in the upper half-plane, $\omega ''>0$. For each of these zeros there exists a solution to Maxwell equations ${{\bf{E}}_{{\rm{zero}}}}(r,t)$ satisfying the incoming boundary conditions with time dependence described by the factor ${e^{ - i{\omega _0}t}}$, i.e., with exponentially growing amplitude ${e^{\omega ''t}}$, such that the scattered fields are absent. In the transient, as the excitation grows, the system looks like a perfectly absorbing structure, despite the fact that it has no material loss. However, the energy is not absorbed, but instead ideally stored inside the open resonant cavity without scattering, since the outgoing fields are forbidden. Since the exponentially growing incident field cannot be maintained indefinitely, when it stops or when a nonlinearity is reached, the stored energy will leak out of the cavity and released.

Similarly to the case of coherent perfect absorption~\cite{Chong2010,Zhang2012, Yoon2012, Ignatov2016}, in order to engage the complex zero, the system needs to be excited by coherent partial waves. For example, a two-port system (such as the dielectric slab in Fig.~\ref{fig1}) should be excited by two coherent exponentially growing signals from the two sides. This excitation allows to control the storage and release of energy in the system without relying on nonlinearity, but simply based on the spatiotemporal coherence of the incoming fields. In order to illustrate this effect, let us first consider the one-dimensional case of a homogeneous dielectric slab of thickness $L$ located in free space, as depicted in Fig.~\ref{fig1}(a). For the sake of simplicity, we focus on the case of normal incidence. The amplitudes of the output fields ${b_{\rm{r}}}$ and ${b_{\rm{l}}}$ are related to the amplitudes of the input fields ${a_{\rm{r}}}$ and ${a_{\rm{l}}}$  via the scattering matrix $S$:
\begin{equation}
\left( {\begin{array}{*{20}{c}}
{{b_{\rm{r}}}}\\
{{b_{\rm{l}}}}
\end{array}} \right) = \left( {\begin{array}{*{20}{c}}
r&t\\
t&r
\end{array}} \right)\left( {\begin{array}{*{20}{c}}
{{a_{\rm{r}}}}\\
{{a_{\rm{l}}}}
\end{array}} \right)
\end{equation}
where $r$ and $t$ are the slab reflection and transmission coefficients, respectively. The above  $S$-matrix has a pair of eigenvalues ${s_ \pm } = r \pm t$, corresponding to symmetric and anti-symmetric incident fields~\cite{Chong2010}. The zeros of these eigenvalues, if occurring at real-valued frequencies, support coherent perfect absorption. For a lossless non-magnetic ($\mu  = 1$) slab with relative refractive index $n=3$, however, the zeros are necessarily confined in the upper half of the complex-frequency plane, as shown in Fig.~\ref{fig1}(b) and, not surprisingly, no perfect absorption can take place upon illumination with a harmonic field.

\begin{figure*}
\centering
\includegraphics[width=1.6\columnwidth]{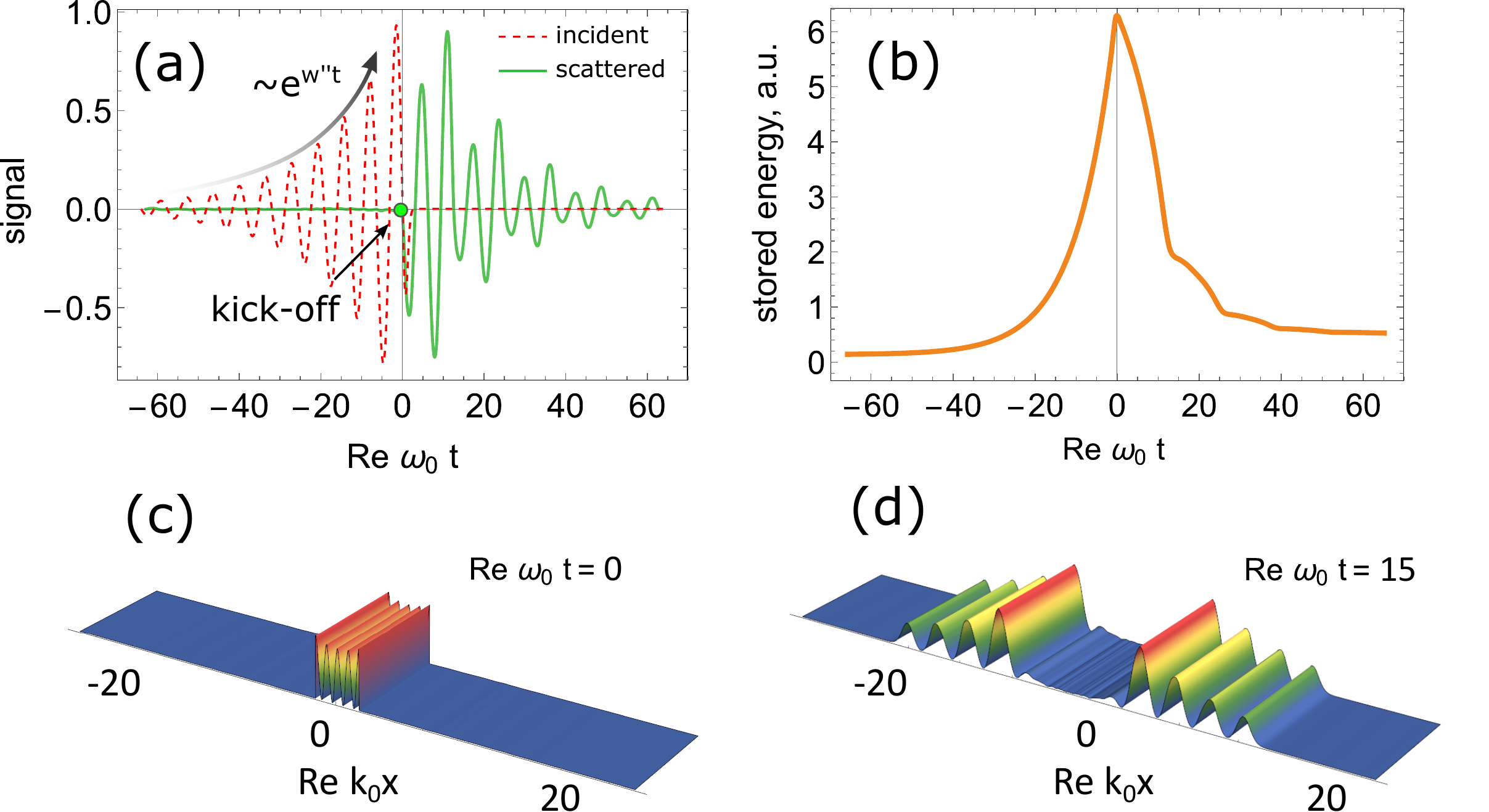}
\caption{Demonstration of coherent virtual absorption in a 1D lossless system. (a) Scattered (solid green curve) field by the planar slab in Fig.~\ref{fig1} upon illumination with exponentially growing field (dashed red curve) matching the complex-frequency scattering zero. Exponential growth stops at $\omega 't = 0$, marking the kick-off for output fields. (b) Electromagnetic field energy stored in the dielectric slab as a function of time for the incident pulse shown in (a). (c,d) Electric field intensity profile (c) exactly at $\omega 't = 0$, and (d) during re-radiation of the stored energy, $\omega 't = 15$.}
\label{fig2}
\end{figure*}

We now tailor the incident field such that during a finite interval of time it mimics the one associated with one of these complex zeros. To do that, we choose an incident pulse with time evolution ${E_{{\rm{inc}}}}(t) = {e^{ - i\omega 't}}{e^{\omega ''t}}\theta ( - t) + \exp ( - i\omega 't) \cdot \exp ( - {(t/{\tau _a})^2}) \cdot \theta (t)$, where $\theta (t)$ is the Heaviside step-function; the second term ensures continuous decay of the incident field after $t=0$. Specifically, we choose a zero of the symmetric eigenvalue located at the complex frequency ${k_0}L \approx 4.2 + 0.2i$, highlighted by a glow in Fig.~\ref{fig1}(b) with $k_0$ being the free-space wavenumber. The real part of the resulting incident field is shown in Fig.~\ref{fig2}(a) for ${\tau _a} = 1/\omega '$, which results in a fast decay of the incident field to zero after the exponential cutoff at  $t = 0$.

Owing to the linearity of the system, we can calculate its response via convolution of the incident field with the response function ${s_ + } = r + t$. The resulting output field amplitude is shown in Fig.~\ref{fig2}(a) by the green solid curve. Interestingly, the output (reflected and transmitted) fields are completely absent at negative times, despite the system being lossless and being driven by the incident field. Because of the absence of dissipation, the incident energy cannot be dissipated and, therefore, it is perfectly stored with unitary efficiency within the resonator without scattering. This energy is released at $\omega 't > 0$, when the incident field stops the exponential growth and can no longer continue pumping the virtual absorbing mode. Notably, the resulting scattered field does not represent the time-reversed incident field, but rather it is a linear superposition of several eigenmodes of the cavity.

The observed behavior can be interpreted as transient destructive interference of the scattered waves. A lossless system necessarily scatters incident harmonic fields due to the unitarity of its $S$ matrix. But, unlike the case of harmonic excitation, we provide a growing amount of energy to the system with the correct amplitude and phase, properly weighed so that this new energy cancels the previously scattered signal. In order to get more insights into the proposed phenomenon of coherent virtual absorption, we simulated the temporal dynamics of the electromagnetic field inside the slab using CST Microwave Studio. The results confirm the absence of scattered fields during the exponential excitation from $\omega 't =  - \infty $ up to $\omega 't = 0$. The temporal dynamics of the electric field energy inside the slab, Fig.~\ref{fig2}(b), clearly demonstrates the accumulation of the incident energy inside the system, and its triggered release on demand when the excitation changes. The field distribution profile for $\omega 't \le 0$, presented in Fig.~\ref{fig2}(c), shows that the energy is concentrated within the slab without any scattering. For the sake of clarity, in the figure we have subtracted the propagating incident field from the total field. At the instant in which we interrupt the exponential pumping ($\omega 't = 0$), the accumulated energy is released through free-space radiation, which, given the time-reversal symmetry of the problem, decays with the same time constant as the excitation. This is illustrated in Fig.~\ref{fig2}(d), showing the electric field distribution profile at $\omega 't = 15$. Similar results may be obtained engaging the complex zeros associated with the anti-symmetric excitation, noting that in both cases the coherence of the impinging fields from the two sides plays a crucial role in controlling the storing and release of energy by the slab. This coherence may be exploited to control in real-time this phenomenon for low-energy memory and switching.

\begin{figure}[!b]
\centering
\includegraphics[width=.7\columnwidth]{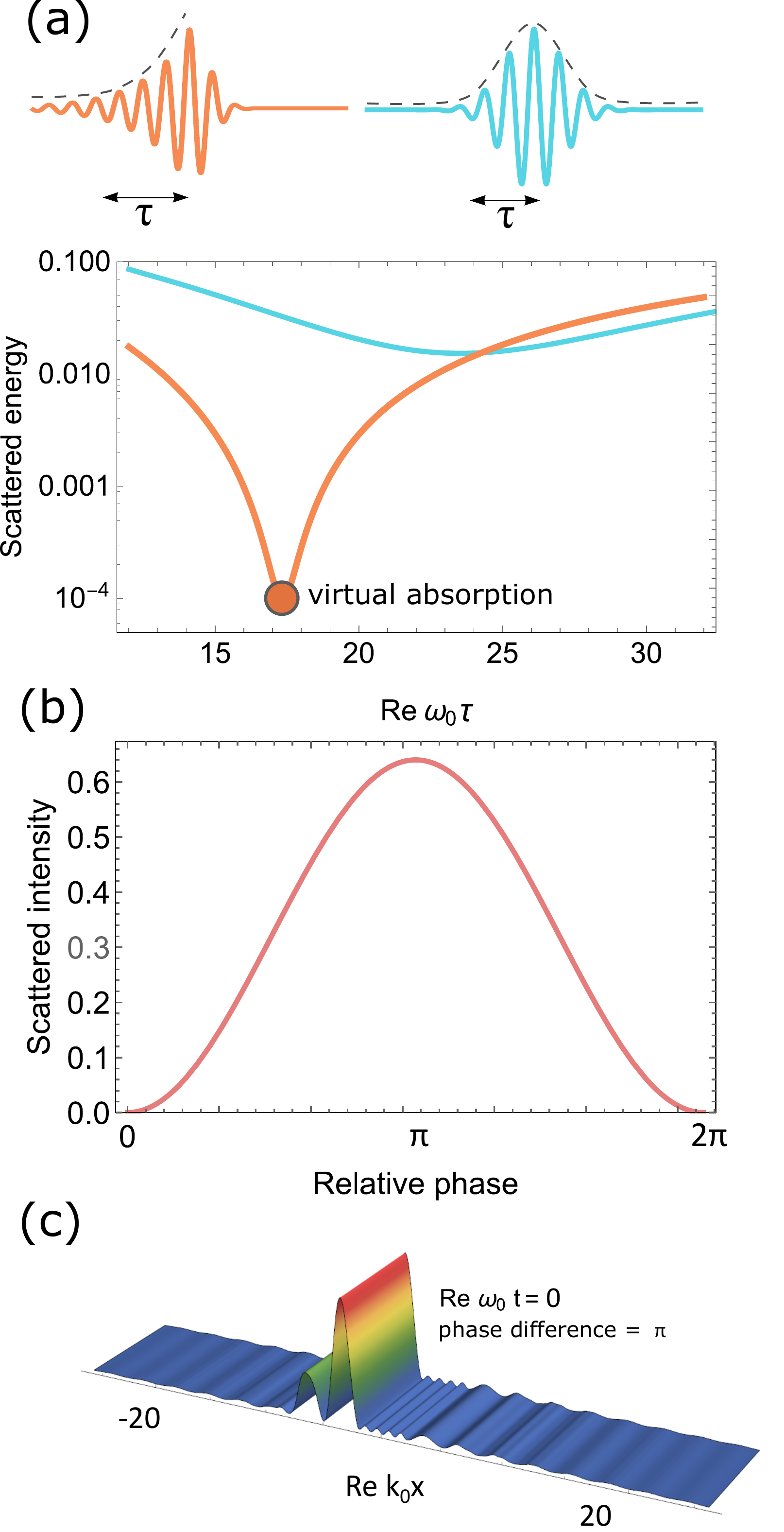}
\caption{(a) Scattering of exponential (orange) and Gaussian (cyan) pulse from the dielectric slab. The plot shows the total scattered power integrated at negative times $\omega 't \le 0$ normalized to the total incident power as a function of the incident pulse time constant. (b) Total scattering (normalized by the instantaneous value of incident intensity) of a virtually absorbing pulse as a function of relative phase between two incident waves. (c) The corresponding electric field intensity profile at $\omega 't = 0$ upon illumination with two waves with $\pi$ phase difference.}
\label{fig3}
\end{figure}

In order to illustrate the role of the complex scattering zero in this unusual response, we simulated the transient dynamics of the same structure illuminated by finite exponentially growing pulses varying the time constants $\tau  = 1/\omega ''$ and by a Gaussian pulse ${E_{\rm{G}}}(t) = \exp ( - i\omega 't) \cdot \exp ( - {(t/\tau )^2})$ for different pulse widths $\tau $. The total back-scattered intensity integrated at negative times $\omega 't \le 0$, shown in Fig.~\ref{fig3}(a), clearly demonstrates that complete suppression of scattering cannot be achieved for the Gaussian input, and matching of the incident field to the scattering zero is essential for coherent virtual absorption. 

Figure~\ref{fig3}(a) highlights the importance of shaping the envelope of the incident field to engage the complex zero of a lossless system. Our analysis implies that the majority of the energy stored in the system arrives during a time interval $\tau  = 1/\omega ''$, which represents the storage and, by reciprocity, the decay time of the structure. However, so far we have considered a rather plain design for the coherent virtual absorber, i.e., a dielectric slab. By tailoring the design of the open system, it may be possible to tailor at will the effective interaction time by controlling the position of the complex zero of interest. For instance, we envision an open lossless resonator whose scattering zero is located close to the real frequency axis, which may be achieved introducing highly reflective mirrors on both surfaces of the dielectric slab, which will increase the $Q$-factor of the resonance and, by time-reversal symmetry, of the associated complex zero, and thus the interaction time $\tau$ by orders of magnitude. Another promising way to achieve an arbitrary tailoring of the time scale associated with the coherent virtual absorption phenomenon is to take advantage of the concept of embedded scattering eigenstates~\cite{Marinica2008,Monticone2014, Lannebere2015,Hsu2016,Gansch2016,Silveirinha}. An embedded eigenstate is a nonradiating eigenmode of an open structure, which is compatible with the continuum of radiation modes but it does not radiate. The proper design of a structure supporting an open embedded eigenstate, may enable the excitation of coherent virtual absorption states with extremely long interaction times, consequently avoiding the need for a fast growing envelope of the excitation fields. 
We stress that an ideal embedded eigenstate is not required  to achieve this effect, as it would completely suppress coupling between incident wave and the eigenmode. Instead, a small imaginary part in the frequency of the complex zero, as expected in any realistic implementation of the phenomenon, would turn an ideal embedded eigenstate into a very high-$Q$ resonance accessible from the outside, and the coherent virtual absorption phenomenon would enable its ideal, efficient excitation.

\begin{figure*}
\centering
\includegraphics[width=1.7\columnwidth]{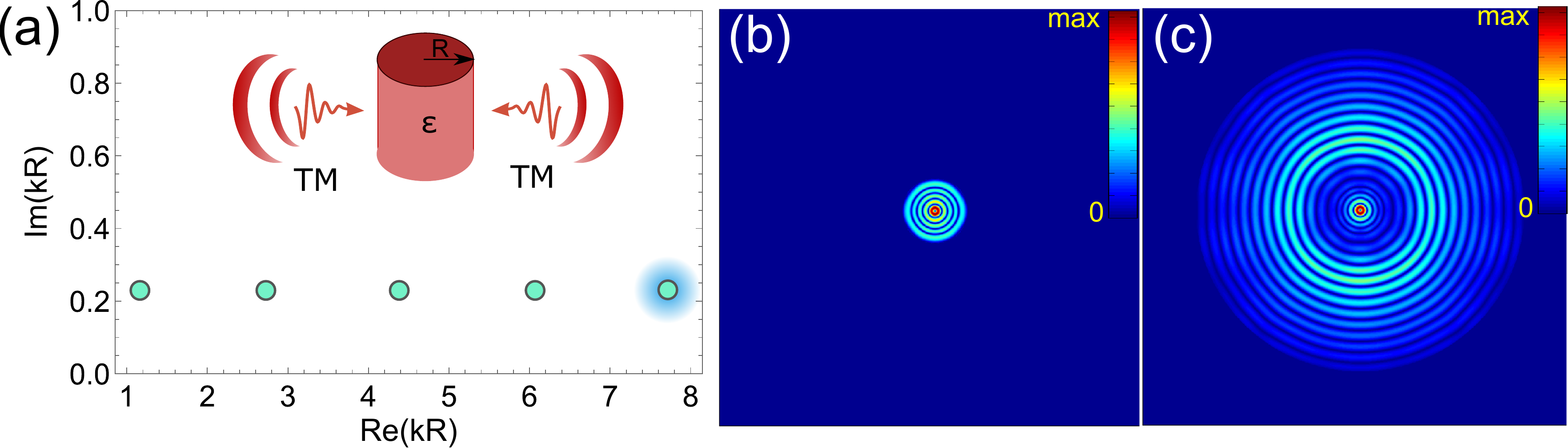}
\caption{(a) Location of the $S$-matrix zeros of a dielectric cylinder with relative refractive index $n=2$ and radius $R$ for TM cylindrical wave excitation. The inset shows the schematic geometry of the problem. (b,c) Electric field intensity profile (b) right before interrupting the exponential growth, $\omega 't = 0$, and (c) during re-radiation of the stored energy at $\omega 't = 20$. The incident fields are subtracted from the total fields for clarity.}
\label{fig4}
\end{figure*}

Although we have employed a two-port system in the above example, incorporation of two coherent beams is, in principle, not necessary to realize virtual absorption. A similar phenomenon can arise in a slab backed by a mirror, with a single plane wave incident from the open side. The scattering eigenvalues for such a boundary value problem yield zeros in the complex frequency plane, based on which we can extract the required temporal shape of the incident wave resulting in scattering-free energy storage inside the slab. Nevertheless, the presence of two input channels, as in the scenario considered in Fig.~\ref{fig1}, provides an important degree of freedom to control the scattering through the coherence of the two incoming waves. Indeed, similarly to CPA, but importantly without relying on Ohmic losses, the response of our system is sensitive to relative phase of the two incident waves. In the case of harmonically oscillating fields losses are necessary for controlling the output intensity~\cite{Chong2010,Zhang2012}, since scattering from a lossless structure is always unitary. In contrast, the use of non-harmonic excitation allows us to modulate the total scattered intensity in a lossless system. This property of the system is illustrated in Fig.~\ref{fig3}(b), in which phase-dependent total scattering (normalized by the instantaneous value of incident intensity) is shown, revealing the possibility of modulating the scattering versus stored energy in a lossless system in real time by varying the relative phase of one of the incoming waves. The electric field intensity distribution taken at the cut-off point $t=0$ for $\pi$ phase difference presented in Fig. \ref{fig3}(c) indeed demonstrates nonzero scattered field outside the slab.

The coherent virtual absorption phenomenon described in this paper can be also generalized to finite size and open scatterers in two and three dimensions. Consider a lossless dielectric non-magnetic cylinder with relative refractive index $n=2$ and radius $R$ located in free-space with the axis collinear with the $z$-axis. We illuminate the structure with a converging cylindrical harmonic, as sketched in Fig. 4. The zeros of the scattering coefficient in the complex frequency plane can be calculated from the scattering amplitude~\cite{Noh2012}
\begin{equation}
s = \frac{{n{J_m}(n{k_0}R)H{{_m^{(2)}}^\prime }({k_0}R) - {n_0}{J_m}^\prime (n{k_0}R)H_m^{(2)}({k_0}R)}}{{{J_m}^\prime (n{k_0}R)H_m^{(1)}({k_0}R) - n{J_m}(n{k_0}R)H{{_m^{(1)}}^\prime }({k_0}R)}},
\end{equation}
where $H_m^{(1)}$ [$H_m^{(2)}$] is the $m$-th order Hankel function of the first [second] kind, ${J_m}$ is the Bessel function of the first kind, and the prime apex denotes the first-order derivative. For excitation with a transverse magnetic (TM) $m=0$ harmonic, the scattering zeros are shown in Fig. 4(a). We focus on the zero at complex frequency ${k_0}R \approx 7.8 + 0.25i$; Figs.~\ref{fig4}(b ,c) demonstrate the electric field intensity profile right before interrupting the exponential growth at $\omega 't = 0$ (b), and during the re-radiation of the stored energy at $\omega 't = 20$ (c). As in the planar scenario, as soon as the exponential pumping is interrupted, the stored energy leaves the dielectric rod through radiation in free space, with reversed temporal dynamics compared to the coherent perfect absorption, while the scattering is totally suppressed throughout the excitation period.

Although we considered a non-dispersive material in the examples above, it is important to highlight that this effect is robust against material dispersion. Indeed, any frequency dispersion of material parameters will lead to a modification of both the real frequencies response \emph{and} spectral positions of poles and zeros. On the other hand, as was pointed out in Refs.~\citenum{Bonod13,Bonod13b}, the position of these poles and zeros is unambiguously linked to the system response at real frequencies. Therefore, the position of zeros entails information about material dispersion, enabling virtual absorption in the same manner as demonstrated above for a non-dispersive material.

Following the same argument, it becomes clear that this effect is not exclusive to lossless systems. Our additional simulations reveal that the effect is absolutely robust to the degree of material loss, provided that the incident pulse has a proper time dependence matching the position of the scattering zero in the complex frequency plane.
In contrast to lossless systems, however, ideal light storage will not take place in lossy systems, since only a fraction of the incident signal will be re-emitted, while a finite amount of energy will be absorbed by the structure during the interaction with the incident pulse; yet, ideally zero scattering will be observed also in such lossy systems. 

While this response can be observed in canonical structures, such as planar slabs and dielectric rods, it may be challenging to verify this phenomenon in optics in a practical setup using conventional structures and materials, owing to the required exponential growth. For the example of Fig.~\ref{fig1}, for instance, the $Q$-factor of the absorbing mode is around 20, which yields a characteristic growth time of about 40 fs for a typical 533 nm laser. Such fast modulation of the incident field may be difficult to realize in practice. However, as noted above, one may employ high-$Q$ resonators and bound states in the continuum in order to lower the required growth rate. Employing a photonic crystal cavity or a microring resonator with $Q$-factors around 10$^9$ (Refs.~\citenum{Vahala,Grudinin}) would result in a growth time constant in the order of 1 ns or longer. Modulation at this speed may be performed with an electrostatically controlled Pockels cell or ultrafast optical modulators~\cite{Switching}. At the same time, excitation of this resonator could be performed via a near-field coupled fiber with the propagating mode amplitude increasing exponentially in time.

Choosing an electromagnetic resonator with high $Q$-factor and small mode volume may enable the proposed effect for subwavelength light focusing. For example, the cylinder depicted in Fig.~\ref{fig4} may localize the light in subwavelength dimensions with unitary efficiency, if excited by a properly shaped incident field. In some sense, the proposed effect can be interpreted as an example of time reversal techniques for electromagnetic field focusing, as proposed in Refs.~\cite{Fink, Fink_a}. However, there is no strict analogy between the time-reversal and complex zero excitation approaches. The time-reversal technique relies on a time reversal mirror that is intended to reverse propagation of an arbitrary emitted signal, making it converge towards the source. The proposed technique, instead, engages a single complex zero of a resonator to ideally capture a certain incident signal in a lossless system without back scattering. It relies on exact knowledge of the cavity eigenmodes, which have to be found either numerically \cite{Powell}, or analytically whenever possible. In addition, this storage can be controlled in real-time by detuning the relative phase of the incoming beams, as discussed above.


Finally, we notice that the described effect is not limited to optics. Indeed, the scattering matrix formalism is equally applicable to lower frequencies, for which slower temporal variations are expected, to acoustics, and to single-particle quantum mechanical problems. Since the Hamiltonians of quantum mechanical systems are naturally Hermitian, such systems have scattering zeros in the upper half-plane of the complex energies. Therefore, properly shaping the temporal wave function of an incident particle~\cite{Warren1993,Weinacht1999} is expected to enable ideal storing of the wave function inside the potential without scattering, just like in the electromagnetic case described here.

\section{Conclusion}
In conclusion, here we have introduced the concept of coherent virtual absorption in a lossless electromagnetic system. This phenomenon arises when the incident electromagnetic field matches the spatio-temporal distribution of a complex scattering zero of a lossless system. The interaction with a properly shaped incident pulse results in ideal storage and accumulation of the incident energy inside the system with zero reflections and back-scattering. Interruption of the exponential driving, or suitable dephasing of the impinging fields, gives rise to the release of energy stored in the lossless system through radiation in the background. The effect is robust against frequency dispersion of the system material, possible dissipation and finite geometry of the structure. Our results highlight a counterintuitive transient response of lossless structures and may open the door to novel ways to dynamically control, store and release light and other waves with extremely low energy.

\begin{acknowledgments}
We acknowledge stimulating discussions with Dr.~Alexander Poddubny and Dr.~Francesco Monticone.
This work was partially supported by the Air Force Office of Scientific Research. D.G.B. acknowledges support from the Knut and Alice Wallenberg Foundation.
\end{acknowledgments}

\bibliography{virtual}
\end{document}